\documentclass{aastex631}

\begin{document}

\title{Background Contamination of the Project Hephaistos Dyson Spheres Candidates}

\author[0000-0003-4988-3513]{Tongtian Ren}
\affiliation{Jodrell Bank Centre for Astrophysics, Department of Physics and Astronomy, School of Natural Sciences, University of Manchester, Oxford Road, Manchester M13 9PL, UK}

\author[0000-0001-6714-9043]{Michael A. Garrett}
\affiliation{Jodrell Bank Centre for Astrophysics, Department of Physics and Astronomy, School of Natural Sciences, University of Manchester, Oxford Road, Manchester M13 9PL, UK}
\affiliation{Leiden Observatory, Leiden University, PO Box 9513, NL-2300 RA Leiden, the Netherlands}
\affiliation{University of Malta, Institute of Space Sciences and Astronomy, Msida, MSD2080, Malta}

\author[0000-0003-2828-7720]{Andrew P.V. Siemion}
\affiliation{Breakthrough Listen, University of California, Berkeley, CA 94720, USA}
\affiliation{Berkeley SETI Research Center, University of California, Berkeley, CA 94720, USA}
\affiliation{SETI Institute, 339 Bernardo Avenue, Suite 200, Mountain View, CA 94043, USA}
\affiliation{Department of Physics, University of Oxford, Denys Wilkinson Building, Keble Road, Oxford OX1 3RH, UK}
\affiliation{Jodrell Bank Centre for Astrophysics, Department of Physics and Astronomy, School of Natural Sciences, University of Manchester, Oxford Road, Manchester M13 9PL, UK}
\affiliation{University of Malta, Institute of Space Sciences and Astronomy, Msida, MSD2080, Malta}



\begin{abstract}

Project Hephaistos recently identified seven M-dwarfs as possible Dyson Spheres (DS) candidates. We have cross-matched three of these candidates (A, B \& G) with radio sources detected in various all-sky surveys. The radio sources are offset from the Gaia stellar positions by $\sim 4.9$, $\sim 0.4$ and $\sim 5.0$ arcseconds for candidates A, B, and G respectively. We propose that DOGs (Dust obscured galaxies) lying close to the line-of-sight of these M-dwarf stars significantly contribute to the measured WISE mid-IR flux densities in the WISE W3 and W4 wavebands. These three stars have therefore been misidentified as DS candidates. We also note that with an areal sky density of $9 \times 10^{-6}$ per square arcsecond, Hot DOGs can probably account for the contamination of all 7 DS candidates drawn from an original sample of 5 million stars. 

\end{abstract}



\section{Introduction} \label{sec:intro}

Project Hephaistos \citep{2024MNRAS.531..695S} recently proposed seven Dyson Sphere (DS) candidates by identifying sources with an IR excess from a sample of 5 million stars detected by Gaia, 2MASS, and WISE. The DS candidates are all M-type dwarfs, and natural explanations such as warm debris disks are ruled out as potential contaminating sources. To understand more about other potential sources of contamination in these systems, we compared the position of these stars with the publicly available data from various all-sky radio surveys. 

\section{Search for radio source counterparts} 
\label{sec:radio}

We cross-matched the seven candidates with the Very Large Array Sky Survey (VLASS, \citet{2021ApJS..255...30G}), Rapid ASKAP Continuum Survey (RACS, \citet{2021PASA...38...58H}), the FIRST survey \citep{2015ApJ...801...26H}, the NRAO VLA Sky Survey (NVSS, \citet{1998AJ....115.1693C}), and the TIFR GMRT Sky Survey (TGSS, \citet{2017A&A...598A..78I}). We searched for radio sources within a radius of 10 arcseconds of the Gaia positions. We found radio sources associated with candidates A, B, and G with offsets of $\sim 4.9$, $\sim 0.4$ and $\sim 5.0$ arcseconds respectively. Candidate G is detected in multiple radio surveys. Table \ref{tab:table1} summarises our findings.  

\begin{table}
    \centering
    \footnotesize 
    \begin{tabular}{cccccccccc}
         \hline
         Candidate  & Survey & ID & RA(J2000) & DEC(J2000) & Total offset & RA offset & DEC offset & Flux density & Frequency \\
         Label & &  & (hms) & ($\arcdeg \arcmin \arcsec$) & (arcsec) & (arcsec) & (arcsec) & (mJy) & (MHz) \\ 
         \hline
         A & RACS-DR1 & J124512.7-265206 & 12 45 12.783 & -26 52 06.204 & 4.880 & -2.215 & -4.348 & 1.75 & 887.5 \\
         B & RACS-DR1 & J035603.8-403148 & 03 56 03.831 & -40 31 48.187 & 0.388 & 0.350 & -0.167 & 2.90 & 887.5\\
         G & VLASS & J233532.86-000424.9 & 23 35 32.865 & -00 04 24.945 & 5.686 & 5.638 & -0.737 & 25.45 & 3000\\
         G & FIRST & J233532.8-000425 & 23 35 32.864 & -00 04 25.300 & 5.728 & 5.623 & -1.092 & 33.59 & 1400 \\
         G & NVSS & J233532-000425 & 23 35 32.780 & -00 04 25.000 & 4.434 & 4.363 & -0.792 & 33.90 & 1400\\
         G & RACS-DR1 & J233532.8-000425 & 23 35 32.849 & -00 04 25.244 & 5.500 & 5.398 & -1.036 & 46.39 & 887.5\\
         G & TGSS & J233532.8-000426  & 23 35 32.849 & -00 04 26.256 & 5.499 & 5.398 & -1.048 & 113.20 & 150\\
         \hline
    \end{tabular}
    \caption{The radio source positions, offsets and flux densities associated with DS candidates A, B and G.}
    \label{tab:table1}
\end{table}

\section{Discussion}
\label{sec:discuss}

\begin{figure}
    \centering
    \includegraphics[width=1\linewidth]{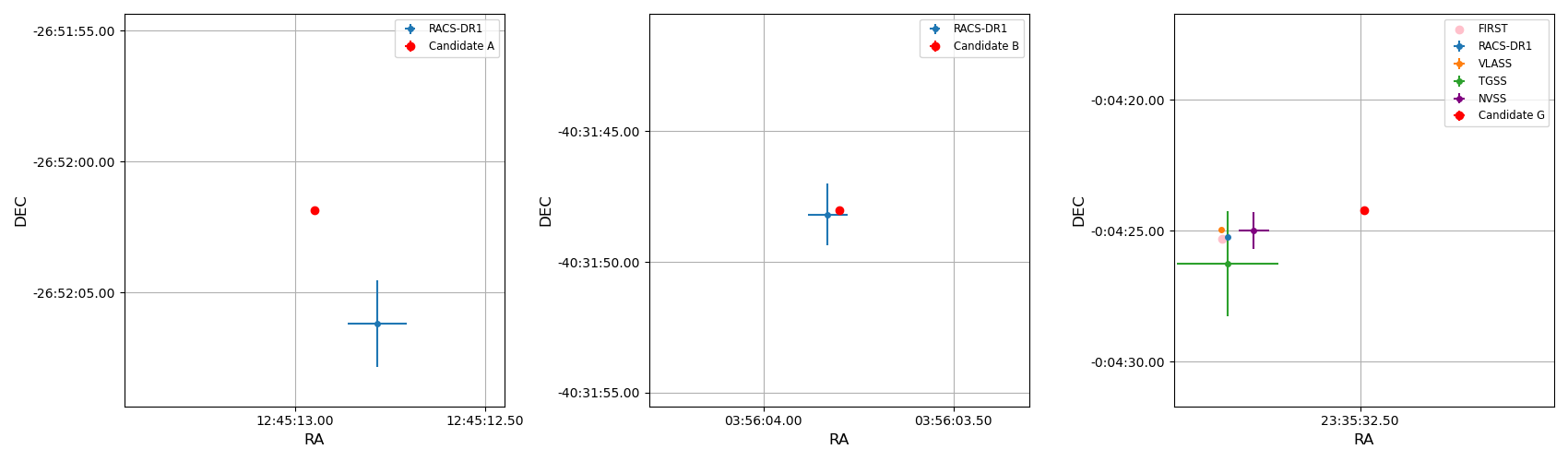}
    \caption{The offset positions of the radio sources associated with DS candidates A, B and G. Each field is 15 x 15 arcseconds in extent. The Gaia coordinates of the DS candidates define the centre of each field. }
    \label{fig:locations}
\end{figure}

Candidates A and G are associated with radio sources offset approximately $\sim 5$ arcseconds from their respective Gaia stellar positions. (see also Fig.\ref{fig:locations}).
We suggest that these radio sources are most likely to be DOGs (dust-obscured galaxies) that contaminate the IR (WISE) Spectral-Energy Distributions (SEDs) of the two DS candidates. The offsets for candidate B are smaller (Fig.\ref{fig:locations}),
approximately $\sim 0.35$ arcsecond. Since M-dwarfs very rarely present persistent radio emission ($\leq 0.5$\% of the sample observed by \citet{2021NatAs...5.1233C}), we suspect that this radio source is also associated with a background DOG lying very close to the line-of-sight. We note that the radio source associated with G has a steep spectral index with a best fit of $\alpha=-0.52 \pm 0.02 $ - this value is typical of synchrotron emission from a radio-loud AGN with extended jets. 

One specific class of background AGN that can explain the observations are Hot DOGs (hot dust-obscured galaxies) \citet{2015ApJ...804...27A}. Hot DOGs have dust temperatures $\geq 60$~K and are detected as WISE W1 and W2 dropouts - they are well detected at longer wavelengths in W3 and W4 \citep{2015ApJ...805...90T}. With a resolution of 6-12 arcseconds across the W1-W4 bands, the radio counterparts of A, B, and G all fall within the primary response of WISE.  

Hot DOGs also have a surface density of approximately 1 per 31 square degrees \citep{2015ApJ...804...27A}, which translates to about $9 \times 10^{-6}$ per square arcsecond. This density is therefore sufficient to explain the levels of contamination observed in large-scale surveys like the one conducted for Project Hephaistos, which analysed approximately 5 million stars. We propose that all seven DS candidates reported by \citet{2024MNRAS.531..695S} 
have very likely been misidentified, with their SEDs being significantly contaminated by background Hot DOGs in the WISE W3 and W4 bands. In this scenario, the other 4 DS candidates (C, D, E, and F) are presumably radio quiet systems. Deeper, and higher-resolution radio observations of the 7 candidates are warranted.


\vspace{5mm}





\bibliography{sample631}{}
\bibliographystyle{aasjournal}



\end{document}